\documentclass[pre,twocolumn,showpacs,preprintnumbers,amsmath,amssymb]{revtex4}

\usepackage{mathrsfs}
\usepackage{graphicx}
\usepackage{dcolumn}
\usepackage{bm}

\begin{document}


\title{Geometrical Expression of Excess Entropy Production}
\author{Takahiro Sagawa$^{1,2}$}
\author{Hisao Hayakawa$^2$}
\affiliation{$^1$The Hakubi Center, The Kyoto University, Yoshida-Ushinomiya-cho,
Sakyo-ku, Kyoto 606-8302, Japan \\
$^2$Yukawa Institute for Theoretical Physics, The Kyoto University, Kitashirakawa Oiwake-Cho, 606-8502 Kyoto, Japan
}
\date{\today}

\begin{abstract}
We derive a geometrical expression of the excess entropy production for quasi-static transitions between nonequilibrium steady states of Markovian jump processes, which can be exactly applied to nonlinear and nonequilibrium situations.  The obtained expression is  geometrical;  the excess entropy production depends only on a trajectory in the parameter space,  analogous to the Berry phase in quantum mechanics.    Our results imply that  vector potentials are needed to construct the thermodynamics of nonequilibrium steady states.
\end{abstract}


\maketitle

\section{Introduction}

Investigating thermodynamic structures of nonequilibrium steady states (NESSs) has been a topic of active researches in nonequilibrium statistical mechanics~\cite{Landauer,Oono,Raulle,Komatsu1,Komatsu2,Saito,Sasa,Seifert,Hatano,Speck,Esposito1,Esposito2,Lebowitz,Nemoto,Komatsu0,Maes1,Maes2}. For example, the extension of the relations in equilibrium thermodynamics, such as the Clausius equality, to NESSs is a great challenge~\cite{Landauer,Oono,Raulle,Komatsu1,Komatsu2,Saito,Sasa,Seifert}.  The extended thermodynamics, which is called steady state thermodynamics (SST)~\cite{Oono}, is expected to be useful to analyze and to predict the dynamical properties of NESSs.  However, the complete picture of SST has not been  understood.

In equilibrium thermodynamics, the Clausius equality tells us how one can determine thermodynamic potential (entropy) by measuring the heat: 
\begin{equation}
\Delta S - \sum_\nu \beta^\nu Q^\nu = 0,
\label{Clausius}
\end{equation}
which is universally valid for quasi-static transitions between equilibrium states.  Here, $\nu$ is an index of the heat baths, $\beta^\nu$ is the inverse temperature of bath $\nu$, $Q^\nu$ is the heat that the system absorbed from bath $\nu$, and $S$ is the Shannon entropy of the system.   The second term of the left-hand side (LHS) of (\ref{Clausius}) is called the entropy production in the baths.
To generalize the Clausius equality to nonequilibrium situations,  it has been proposed~\cite{Landauer,Oono} that heat $Q^\nu$ needs to be replaced by excess heat $Q_{\rm ex}^\nu$,  which describes an additional heat induced by a transition between NESSs with time-dependent external control parameters such as the electric field.  Correspondingly,  the total heat can be decomposed as $Q^\nu = Q_{\rm ex}^\nu + Q_{\rm hk}^\nu$, where housekeeping heat $Q_{\rm hk}^\nu$ describes the steady heat current in a NESS without any parameter change.  
Quantitative definitions of these quantities will be given later.
One may then expect that there exists some thermodynamic potential $S_{\rm SST}$ which characterizes NESSs such that
\begin{equation}
 \Delta S_{\rm SST} - \sum_\nu \beta^\nu Q^{\nu}_{\rm ex} = 0,
\label{Clausius1}
\end{equation}
holds for quasi-static transitions between NESSs, where the second term of the LHS corresponds to the excess part of the entropy production in the baths.  Komatsu, Nakagawa, Sasa, and Tasaki (KNST) found that $S_{\rm SST}$ in Eq.~(\ref{Clausius1}) is a symmetrized version of the Shannon entropy in the lowest order of nonequilibriumness~\cite{Komatsu1,Komatsu2}.  However, the full order expression of the  extended Clausius equality~(\ref{Clausius1}) has been elusive.   Then, some fundamental questions  arise:  What is the nonequilibrium thermodynamic potential $S_{\rm SST}$ in Eq.~(\ref{Clausius1})  in the full order expression? Does there exist such a potential at all?

In this paper, we answer these questions, and derive a full order expression of the excess entropy production for Markovian jump processes.   We note that driven lattice gases are special cases of our formulation.  We have found that an extended Clausius equality in the form of (\ref{Clausius1}) does not hold in general; scalar thermodynamic potential $S_{\rm SST}$ should be replaced by a  vector potential.   In other words, the first term of the LHS of (\ref{Clausius1}) should be replaced by  a geometrical quantity that depends only on trajectories in the parameter space.  Our result includes  equilibrium Clausius equality~(\ref{Clausius}) and the KNST's extended Clausius equality as special cases.   We will also derive the general condition that there exists a thermodynamic potential $S_{\rm SST}$ such that Eq.~(\ref{Clausius1}) holds.

We have used the technique of the full counting statistics~\cite{Esposito1,Bagrets} to prove our main results.  In the context of the full counting statistics (and also stochastic ratchets), it has been reported~\cite{Parrondo,Sinitsyn1,Sinitsyn2,Sinitsyn3,Jarzynski1,Ohkubo1,Ren,Ohkubo2} that several phenomena in classical stochastic processes are analogous to the Berry's geometrical phase in quantum mechanics~\cite{Berry,Samuel}.  In this analogy, the above-mentioned vector potential corresponds to the gauge field that induces the Berry phase.   Our result can also be regarded as a generalization of these previous studies on the classical Berry phase.

This paper is organized as follows.  In Sec.~II, we formulate the model of our system and define the decomposition of the entropy production into the housekeeping and excess parts based on the full counting statistics.  In Sec.~III, we derive our main results, which consist of the geometrical expressions of the excess parts of the cumulant generating function and the average of the entropy production.  In Sec.~IV, we apply our main results to two special cases; one is equilibrium thermodynamics with the detailed balance, and the other is the KNST's extended Clausius equality.  In Sec.~V, we discuss a quantum dot as a simple example, where  Eq.~(\ref{Clausius1}) does not hold in general.   In Sec.~VI, we conclude this paper with some discussions.

\section{Setup}

We first formulate our setup and define the decomposition of the cumulant generating function of the entropy production into the excess and housekeeping parts.

\subsection{Dynamics}

We consider Markovian jump processes with $N < \infty$ microscopic states.  Let $p_x$ be the probability that the system is in state $x$.  The probability distribution of the system is then characterized by vector $| p \rangle :=  [p_1, p_2, \cdots, p_N]^T$, where $1, 2, \cdots, N$ describe the  states, and ``$T$'' describes the transpose of the vector.
The time evolution of the probability distribution is given by a master equation $| \dot p (t) \rangle = R (\bm{\alpha} (t)) | p (t) \rangle$,
where $| \dot  p (t) \rangle$ describes the time derivative of $| p (t) \rangle$, $R(\bm{\alpha})$ is a $N \times N$ matrix characterizing the transition rate of the dynamics with external parameters $\bm{\alpha}$.  Here, the external parameters correspond to, for example, a potential or a nonconservative force applied to a lattice gas, or the temperatures of the heat baths.  We drive the system by changing $\bm{\alpha}$.  For simplicity of notations,  we will often omit   ``$(\bm{\alpha}(t))$'' or  ``$(t)$'' in the following discussions.
We note that $\sum_{x} R_{xy} = 0$ holds for every $y$, where $R_{xy}$ is the $xy$-component of $R$ that characterizes the transition rate from state $y$ to $x$.
We assume that $R$ is irreducible such that $R$ has eigenvalue $0$ without degeneracy  due to the Perron-Frobenius theorem.  We write as $\langle 1 |$ and $| p^S \rangle$ the left and right eigenvectors of $R$ corresponding to eigenvalue $0$ such that $\langle 1 | R = 0$ and $R | p^S \rangle = 0$ hold.  We note that $\langle 1 | = [1, 1, \cdots, 1]$ holds and that $| p^S \rangle = [p^S_1, p^S_2, \cdots, p^S_N]^T$ is the unique steady distribution of the dynamics with a given $\bm \alpha$.  For simplicity, we assume that $R$ is diagonalizable.
We also assume that the transition matrix can be decomposed into the contributions from multiple heat baths, labeled by $\nu$, as $R_{xy} = \sum_\nu R^\nu_{xy}$.

We next introduce the entropy production that depends on trajectories of the system.  Such a trajectory-dependent entropy production has been studied in terms of nonequilibrium thermodynamics of stochastic systems~\cite{Lebowitz,Crooks,Jarzynski2,Seifert}.
The entropy production in bath $\nu$ with transition from $y$ to $x$ is given by
\begin{eqnarray}
\sigma^\nu_{xy} =  \left\{
\begin{array}{l}
\ln \frac{R^\nu_{xy}}{R^\nu_{y x}} = - \beta^\nu Q^\nu_{xy} \  ({\rm if} \ R^\nu_{xy}\neq 0 \ {\rm and} \ R^\nu_{y x}\neq 0),  \\
0 \  ({\rm if} \ R^\nu_{xy} = 0 \ {\rm and}  \ R^\nu_{y x} = 0), 
\end{array}
\right.
\label{entropy}
\end{eqnarray}
where $Q_{xy}^\nu$ is the heat that is absorbed in the system from bath $\nu$ during the transition from $y$ to $x$.  Equality (\ref{entropy}) is consistent with the detailed fluctuation theorem~\cite{Lebowitz,Crooks,Jarzynski2,Seifert}. 
The integrated entropy production from time $0$ to $\tau$ is determined by the trajectory of system's states during the time interval as 
\begin{equation}
\sigma = \sum_{t : \ {\rm jump}} \sigma^\nu_{x(t+0)y(t-0)},
\end{equation}
where the sum is taken over all times at which the system jumps, and $y(t-0)$ and $x(t+0)$ are  the states immediately before and after the jump at $t$, respectively.  
We note that the ensemble average of $\sigma$ is equivalent to the entropy production in the conventional thermodynamics of macroscopic systems. A reason why we consider the trajectory-dependent entropy production lies in the fact that the entropy production is connected to the heat through Eq.~(\ref{entropy}) at the level of each trajectories.

\subsection{Full Counting Statistics}

We then discuss the full counting statistics of $\sigma$.  Let $P(\sigma)$ be the probability of $\sigma$.  Its cumulant generating function is given by 
\begin{equation}
S(i\chi) := \ln \int d\sigma e^{i\chi \sigma}P(\sigma),
\end{equation}
where $\chi \in \mathbb R$ is the counting field.  $S(i \chi)$ leads to the cumulants of $\sigma$ like $\langle \sigma \rangle = \partial S (i \chi) / \partial (i \chi) |_{\chi = 0}$,
where $\langle \cdots \rangle$ describes the statistical average.
To calculate $S(i\chi)$, we define matrix $R_\chi$ as $( R_\chi )_{xy} := \sum_\nu R_{xy}^\nu \exp (i\chi \sigma_{xy}^\nu)$,
and consider the time evolution of vector $| p_\chi (t) \rangle$ corresponding to
\begin{equation}
 | \dot p_\chi (t) \rangle = R_\chi (\bm \alpha (t)) | p_\chi (t) \rangle
\label{master_chi}
\end{equation}
with initial condition $| p_\chi (0) \rangle := | p(0) \rangle$.  The formal solution of Eq.~(\ref{master_chi}) is given by $| p_\chi (t) \rangle = {\rm T} \! \exp_{\leftarrow} \left( \int_0^\tau R_\chi (\bm{\alpha} (t)) dt \right) | p(0) \rangle$, where ${\rm T} \! \exp_{\leftarrow}$ describes the left-time-ordered exponential.
Then we can show that
\begin{equation}
e^{S(i\chi)} = \langle 1 | p_\chi (\tau) \rangle
\end{equation}
holds, where $\langle \cdot | \cdot \rangle$ means the inner product of left and right vectors.

We write the eigenvalues of $R_\chi$  as $\lambda_\chi^n$'s, where  $n=0$ corresponds to the eigenvalue with the maximum real part. 
If $| \chi |$  is sufficiently small, $\lambda_\chi^0$  is not degenerated and $R_\chi$ is diagonalizable.  
We write as $\langle \lambda_\chi^n |$ and $| \lambda_\chi^n \rangle$ the left and right eigenvectors corresponding to $\lambda_\chi^n$,  which we can normalize as $\langle \lambda_\chi^n | \lambda_\chi^m \rangle = \delta_{nm}$  with $\delta_{nm}$ being  the Kronecker's delta.  
In particular, we write $\langle \lambda_\chi^0 | =: \langle 1_\chi |$ and $| \lambda_\chi^0 \rangle =: | p^S_\chi \rangle$.
We note that, if $\chi = 0$,  $\langle 1_\chi |$ and $| p_\chi^S \rangle$ reduce to $\langle 1 |$ and $| p^S \rangle$, respectively.   

\subsection{Decomposition of the Entropy Production}

It is known that $\lambda_\chi^0 (\bm \alpha)$ is the cumulant generating function of $\sigma$ in the steady distribution with parameter $\bm \alpha$.
More precisely, $\lambda_\chi^0 (\bm \alpha)$ satisfies
\begin{equation}
\lambda_\chi^0 (\bm \alpha) = \lim_{\tau \to +\infty} \frac{S(i \chi; \bm{\alpha}; \tau)}{\tau},
\label{hk_average}
\end{equation}
where $S(i \chi; \bm{\alpha}, \tau)$ is the cumulant generating function of $\sigma$ from $0$ to $\tau$ with $\bm \alpha$ being fixed.

We then decompose the cumulant generating function into two parts:
\begin{equation}
S(i\chi) = S_{\rm hk} (i \chi) + S_{\rm ex} (i \chi),
\end{equation}
where $S_{\rm hk} (i\chi)$ is the house-keeping part defined as
\begin{equation}
S_{\rm hk} (i\chi) : = \int_0^\tau \lambda_\chi^0 (\bm{\alpha} (t)) dt, 
\end{equation}
and $S_{\rm ex} (i\chi)$ is the excess part   defined as $S_{\rm ex}(i\chi) := S (i \chi) - S_{\rm hk} (i \chi)$.  
The average of the excess entropy production is given by 
\begin{equation}
\langle \sigma \rangle_{\rm ex} = \frac{\partial S_{\rm ex} (i \chi)}{\partial (i \chi)} \biggr|_{\chi = 0}.
\label{excess_average}
\end{equation}

We note that the above decomposition is consistent with that in Refs.~\cite{Komatsu1,Komatsu2}.  In fact, from Eqs.~(\ref{hk_average})  and (\ref{excess_average}), we can show 
\begin{equation}
\langle \sigma \rangle_{\rm ex} = \langle \sigma \rangle - \int_0^\tau \langle \dot \sigma  \rangle_{{\rm hk}; \bm{\alpha} (t)} dt,
\end{equation}
where $\langle \dot \sigma \rangle_{{\rm hk}; \bm{\alpha}} := \partial \lambda_\chi^0 (\bm \alpha) / \partial (i \chi) |_{\chi = 0}$ is the long-time average of the entropy production per unit time with $\bm{\alpha}$ being fixed.

\section{Main Results}

We now discuss the main results of this paper, which we will refer to as  Eqs.~(\ref{berry2}) and (\ref{average_e}).  
First of all, we expand $| p_\chi (t) \rangle$ as 
\begin{equation}\
| p_\chi (t) \rangle = \sum_n c_n(t) e^{\Lambda_\chi^n (t)} | \lambda_\chi^n (\bm{\alpha} (t)) \rangle,
\label{c_0}
\end{equation}
where $\Lambda_\chi^n (t) := \int_0^t \lambda_\chi^n (\bm{\alpha}(t')) dt'$.  We can show that $\dot{c}_0 = - \sum_n c_n \langle 1_\chi | \dot{\lambda}_\chi^n \rangle e^{\Lambda_\chi^n - \Lambda_\chi^0}$ and  $\langle 1_\chi | \dot \lambda_\chi^n \rangle = \langle 1_\chi | \dot R_\chi | \lambda_\chi^n \rangle / (\lambda^n_\chi - \lambda_\chi^0)$ hold.  Therefore, if the speed of the change of the external parameters is much smaller than the relaxation speed of the system, we obtain
\begin{equation}
\dot c_0 (t) \simeq - c_0 (t) \langle 1_\chi (\bm{\alpha} (t))  | \dot p_\chi^S (\bm{\alpha} (t)) \rangle.
\label{equation}
\end{equation}
Here, we have  used that the real part of $\Lambda^n_\chi - \Lambda_\chi^0$ is negative for all $n \neq 0$.  
We note that this result is similar (but not equivalent) to the adiabatic theorem in quantum mechanics.

Assume that we quasi-statically change parameter $\bm \alpha$  between time $0$ and $\tau$ along a curve $C$ in the parameter space.   The solution of Eq.~(\ref{equation}) is  given by
\begin{equation}
\begin{split}
c_0 (\tau) &=  c_0 (0) e^{- \int_0^\tau dt  \langle 1_\chi (\bm{\alpha} (t)) | \dot p_\chi^S (\bm{\alpha} (t)) \rangle } \\
&= c_0 (0) e^{- \int_C \langle 1_\chi | d | p_\chi^S \rangle },
\end{split}
\label{berry1}
\end{equation}
where  ``$d$'' on the right-hand side (RHS) means the total differential in terms of $\bm{\alpha}$ such that $d|p^S_\chi \rangle := d\bm{\alpha} \cdot \frac{\partial}{\partial \bm{\alpha}}|p^S_\chi \rangle$.
Let the initial distribution be the steady distribution $| p (0) \rangle = | p^S (\bm \alpha (0)) \rangle$, which leads to $c_0 (0) = \langle 1_\chi (\bm{\alpha} (0)) | p^S(\bm{\alpha}(0)) \rangle$. We then obtain the excess part of the cumulant generating function as
\begin{equation}
\begin{split}
S_{\rm ex} (i\chi) =   &\int_C \langle 1_\chi | d | p_\chi^S \rangle \\
&+ \ln \langle 1_\chi (\bm{\alpha}(0) )| p^S(\bm{\alpha}(0)) \rangle + \ln \langle 1 | p_\chi^S (\bm{\alpha}(\tau)) \rangle,
\end{split}
\label{berry2}
\end{equation}
where the RHS  is geometrical and analogous to the Berry phase in quantum mechanics~\cite{Berry}; it only depends on trajectory $C$ in the parameter space.   More precisely, the  RHS of (\ref{berry2}) is analogous to the non-cyclic Berry phase~\cite{Ohkubo2,Samuel}.
We note that $\Lambda_\chi^n (\tau)$ is analogous to the dynamical phase.   In this analogy, $| p_\chi^S \rangle$ and $R_\chi$ respectively correspond to a state vector and  a Hamiltonian.
Equality~(\ref{berry2}) is our first main result.

In terminologies of the Berry phase,  $\langle 1_\chi | d |  p_\chi^S \rangle$ corresponds to a vector potential or a gauge field whose base space  is the parameter space.  
The second and the third terms of the RHS of (\ref{berry2}) confirms the gauge invariance of $S_{\rm ex}(i \chi)$ as is the case for quantum mechanics~\cite{Samuel}, where the gauge transformation corresponds to the transformation of the left and right eigenvectors of $R_\chi (\bm{\alpha})$ as $\langle 1_\chi (\bm{\alpha}) |  \mapsto \langle 1_\chi (\bm{\alpha}) |  e^{ - \theta (\bm{\alpha})}$ and $| p_\chi^S (\bm{\alpha}) \rangle \mapsto e^{\theta (\bm{\alpha})}| p^S (\bm{\alpha}) \rangle$ with $\theta (\bm {\alpha})$ being a scalar.  
We note that several formulae that are similar to Eq.~(\ref{berry2}) have been obtained for different setups~\cite{Sinitsyn1,Sinitsyn2,Sinitsyn3,Ohkubo1,Ren,Ohkubo2}.

By differentiating Eq.~(\ref{berry2}) in terms of $i \chi$, we obtain a simple expression of the average of the excess entropy production:
\begin{equation}
 \int_C \langle 1' | d | p^S \rangle + \langle \sigma \rangle_{\rm ex} = 0,
\label{average_e}
\end{equation}
where $\langle 1' | := \partial  \langle 1_\chi | /  \partial (i\chi) |_{\chi = 0}$.  
Equality~(\ref{average_e}) is the second main result, which is the full order expression of the average of the excess entropy production.  On the contrary to Eq.~(\ref{Clausius1}), the first term of the LHS of (\ref{average_e}) is not given by the difference of a scalar potential $S_{\rm SST}$, but by a geometrical  quantity. We also refer to $\langle 1' | d | p^S \rangle$ as a vector potential.

We can explicitly calculate $\langle 1' |$.  By differentiate the both-hand sides of $\langle 1_\chi | R_\chi = \lambda_\chi^0 R_\chi$ in terms of $i\chi$, we have $\langle 1' | = - \langle 1 | \partial R_\chi / \partial (i\chi) |_{\chi = 0} R^\dagger + k \langle 1 |$,
where $R^\dagger$ is the Moore-Penrose pseudo-inverse of $R$ and $k$ is an unimportant constant. Therefore, we obtain
\begin{equation}
\langle \sigma \rangle_{\rm ex} = \int_C \sum_{\nu x y z} \sigma_{xy}^\nu R_{xy}^\nu R_{yz}^\dagger dp_z^S.
\end{equation}
Some similar formulae for particle currents have been obtained in Refs.~\cite{Parrondo,Jarzynski1}.

We next consider the condition for the existence of thermodynamic potential $S_{\rm SST}$ that satisfies Eq.~(\ref{Clausius1}). For simplicity, we assume that the parameter space is simply-connected, i.e., there is no ``hole'' or singularity.   The necessary and sufficient condition for the existence of $S_{\rm SST}$ is that the integral in the first term of the LHS of (\ref{average_e}) is always determined only by the initial and final points of $C$; or equivalently, $\oint_C \langle 1' | d | p^S \rangle = 0$ holds for every closed curve $C$.  On the other hand, the Stokes theorem states that $\oint_C \langle 1' | d | p^S \rangle  = \int_S d\left( \langle 1' | d | p^S \rangle \right)$ holds,
where $S$ is a surface whose boundary is $C$, and ``$d$'' means the exterior derivative.  By using the wedge product ``$\wedge$,'' we have $d\left( \langle 1' | d | p^S \rangle \right) = d\langle 1' | \wedge d | p^S \rangle := \sum_x  d 1'_x \wedge d  p^S_x = \sum_{xkl} \frac{\partial  1'_x}{\partial \alpha_k} \frac{\partial p^S_x}{\partial \alpha_l} d\alpha_k \wedge d\alpha_l$,  where $1'_x$ means the $x$-component of vector $\langle 1' |$, and $\alpha_k$ is the $k$-component of $\bm{\alpha}$.   Therefore the necessary and sufficient condition is that 
\begin{equation}
d\langle 1' | \wedge d | p^S \rangle = 0
\label{no_curvature}
\end{equation}
holds in every point of the parameter space.  Equation~(\ref{no_curvature}) is equivalent to 
\begin{equation}
\sum_x \left( \frac{\partial 1' _x}{\partial \alpha_k} \frac{\partial  p^S_x}{\partial \alpha_l} - \frac{\partial  1'_x}{\partial \alpha_l} \frac{\partial p^S_x}{\partial \alpha_k}  \right) = 0
\end{equation}
for all $(k, l)$. In terminology of the gauge theory,   $d\langle 1' | \wedge d | p^S \rangle$ corresponds to the strength of the gauge field or the curvature.  For the case of the $U(1)$-gauge theory, the curvature is the magnetic field.

In  equilibrium thermodynamics, Eq.~(\ref{no_curvature}) holds due to the Maxwell relation, and $\langle 1' | d |  p^S \rangle$ becomes the total differential of the Shannon entropy as we will see in the next section.  On the other hand, Eq.~(\ref{no_curvature}) does not hold for transitions between NESSs in general.  In this sense, vector potential  $\langle 1' | d |  p^S \rangle$ plays a fundamental role  instead of  the scalar thermodynamic potential (i.e., the Shannon entropy) in SST.

\section{Special Cases}

In this section, we discuss two special cases, in which the first term of  the LHS of~(\ref{average_e}) reduces to the total differential of a scalar thermodynamic potential.

\subsection{Equilibrium Thermodynamics}

In general, we can explicitly show that Eq.~(\ref{average_e}) reduces to the equilibrium Clausius equality if the detailed balance is satisfied.   Let $E_x$ be the energy of state $x$. The transition rate is given by $R_{xy} = e^{\beta (E_y - W_{xy})}$ with $W_{xy} = W_{y x}$,  the steady distribution by $p_x^S = e^{-\beta E_x} / Z$ with $Z$ being the partition function, and the entropy production in a bath by  $\sigma_{xy}  = -\beta (E_x - E_y) = -\beta Q_{xy}$.  In the quasi-static limit, the system is in contact with a single heat bath with inverse temperature $\beta$ at each time, while $\beta$ can be time-dependent.  We then obtain 
\begin{equation}
 \langle 1' | d | p^S \rangle = \sum_x \beta E_x dp^S_x = d \left(- \sum_x p_x^S \ln p_x^S \right),
\end{equation}
which means that $\langle 1' | d | p^S \rangle $ is the total differential of the Shannon entropy.   

\subsection{KNST's Extended Clausius Equality}

We now show that Eq.~(\ref{average_e}) reduces to the KNST's  extended Clausius equality~\cite{Komatsu1,Komatsu2} in the lowest order  of nonequilibriumness.  Here we assume that, for every $(x,y)$, there exists at most single $\nu$ that satisfies $R_{xy}^\nu \neq 0$, so that we can remove index $\nu$. This is the same assumption as in Refs.~\cite{Komatsu1,Komatsu2}.
Moreover, we formally introduce the time-reversal of states;  the time-reversal of state $x$, denoted as $x^\ast$, is assigned in the phase space.  Since we are considering stochastic jump processes that do not have any momentum term usually, we just interpret the correspondence $x \mapsto x^\ast$ as a formal mathematical map.   
Correspondingly, we should replace $\ln (R^\nu_{xy} / R^\nu_{y x})$ in Eq.~(\ref{entropy}) by $\ln (R^\nu_{xy} / R^\nu_{y^\ast x^\ast})$.  Only with this replacement, all of the foregoing arguments remain unchanged in the presence of the time-reversal. We also assume that, in thermal equilibrium, $p_{x}^S = p_{x^\ast}^S$ holds.
We define
\begin{equation}
\eta := \sum_{xyz} \ln (p_{x^\ast}^S / p_{y^\ast}^S )R_{xy}R_{yz}^\dagger dp_z^S = \sum_x \left( \ln p_{x^\ast}^S \right) dp_x^S
\end{equation}
and  $\tilde{R}_{xy} := R_{y^\ast x^\ast} p_{x^\ast}^S / p_{y^\ast}^S$.
Here, $\tilde R$ is the  adjoint of  $R$ for the cases of $x = x^\ast$~\cite{Hatano,Esposito2}.  We note that $\sum_x \tilde{R}_{xy} = 0$ holds for every $y$.  Since $R = \tilde R$ holds if the detailed balance is satisfied, we characterize the nonequilibriumness of the dynamics by $\varepsilon := \max_{xy} | (\tilde{R}_{xy} - R_{xy})/R_{xy} |$.
We then obtain 
\begin{equation}
\begin{split}
\langle 1' | d | p^S \rangle +\eta &= \sum_{xyz} \ln (\tilde{R}_{xy} / R_{xy}) R_{xy} R_{yz}^\dagger dp_z^S \\
 &= \sum_{xyz}  (\tilde{R}_{xy} - R_{xy})  R_{yz}^\dagger dp_z^S + O(\varepsilon^2 \Delta) \\
&= O(\varepsilon^2 \Delta),
 \end{split}
 \label{eta1}
 \end{equation}
where $\Delta := \max_x | dp_x^S |$ characterizes the amount of the infinitesimal change of the steady distribution.  
On the other hand, 
\begin{equation}
\eta = d \left( \sum_x p_x^S \ln \sqrt{p_x^S p_{x^\ast}^S} \right) +  O(\varepsilon^2 \Delta)
\label{eta2}
\end{equation}
holds~\cite{Komatsu2}.  From Eqs.~(\ref{eta1}) and (\ref{eta2}),  we obtain
\begin{equation}
\langle 1' | d | p^S \rangle =  d \left( - \sum_x p_x^S \ln \sqrt{p_x^S p_{x^\ast}^S} \right) +  O(\varepsilon^2 \Delta),
\end{equation}
which implies the KNST's extended Clausius equality, where the first term of the RHS is the total differential of the symmetrized Shannon entropy. 
We note that, if we gradually change parameter $\bm \alpha$ from an equilibrium distribution, then the KNST's extended Clausius equality is valid up to the order of $O(\varepsilon^2)$ because $\Delta = O(\varepsilon)$ holds.

\section{Example} 

As a simple example that illustrates the absence of a scalar thermodynamic potential, we consider a stochastic model of a quantum dot that is in contact with two baths that are labeled by  $\nu = L$ and $R$ (see also Fig.~1 (a))~\cite{Bagrets}.  
This model describes the stochastic dynamics of the number of  electrons in the dot by a classical master equation.

An electron is transfered from the baths to the dot  one by one or vise-versa.   We assume that the states of the dot are $x=0$ and $1$, which respectively describe that the electron is absent and occupies the dot.  The probability distribution is described by $| p \rangle = [p_0, p_1]^T$, and the transition rate is given by $R = \sum_{\nu = L, R} R^\nu$ with
\begin{eqnarray}
R^\nu = \left[ 
\begin{array}{cc}
- \gamma_\nu f_\nu & \gamma_\nu (1 - f_\nu)   \\
\gamma_\nu f_\nu & -\gamma_\nu (1 - f_\nu) \\
\end{array} 
\right],
\end{eqnarray}
where $\gamma_\nu$ is the tunneling rate between the dot and bath $\nu$,  and $f_\nu = (e^{\beta (E - \mu_\nu)}+1)^{-1}$ is the Fermi distribution function with $\beta$ being the inverse temperature of the baths,  $\mu_\nu$ being the chemical potential of bath $\nu$, and $E$ being the excitation energy of the dot.  The entropy production is given by $\sigma_{00}^\nu = \sigma_{11}^\nu = 0$ and $\sigma_{10}^\nu = - \sigma_{01}^\nu = \sigma_\nu$ with $\sigma_\nu := \beta (\mu_\nu - E)$.  For simplicity, we set $\gamma_L = \gamma_R  =: \gamma$.  Without loss of generality, we assume that the control parameters are  $\sigma_L$ and $\sigma_R$.  
\begin{figure}[htbp]
 \begin{center}
 \includegraphics[width=80mm]{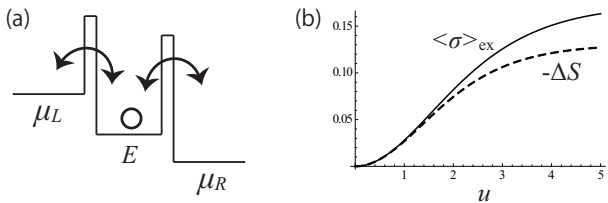}
 \end{center}
 \caption{(a) A schematic of the model of a quantum dot.  A single electron is transfered to/from the two heat baths with chemical potentials $\mu_L$ and $\mu_R$.  (b) $\langle \sigma \rangle_{\rm ex}$ (the solid line) and $-\Delta S$ (the dashed line) for quasi-static processes.  They are coincident with each other up to the second order of the nonequilibriumness of the final state that is denoted by $u$.} 
\end{figure}

We can explicitly calculate the vector potential  as
\begin{equation}
\langle 1' | d | p^S \rangle = -\frac{1}{4} (\sigma_L + \sigma_R) (f_L(1-f_L) d \sigma_L + f_R (1-f_R)d\sigma_R),
\end{equation}
and the curvature as
\begin{equation}
d\langle 1' | \wedge d | p^S \rangle = \frac{1}{4} (f_L (1 - f_L) - f_R (1-f_R)) d\sigma_L \wedge d \sigma_R.
\end{equation}
Therefore, the curvature vanishes only if  $\mu_L = \mu_R$ or $2E = \mu_L + \mu_R$ holds.  The former case corresponds to equilibrium thermodynamics.  Since the curvature vanishes only on the two lines in the two-dimensional parameter space, any scalar potential cannot be defined on the entire parameter space.  We note that the quantities that we have calculated  here are different from those in the previous researches~\cite{Sinitsyn1,Sinitsyn2,Sinitsyn3,Ohkubo1,Ren}.  

As a simple illustration, we consider the following situation.  The dot is initially in thermal equilibrium with $\sigma_L = \sigma_R = 0$. We then quasi-statically change $\sigma_L$ from $0$ to $u$, while $\sigma_R$ is not changed.  We calculate $\langle \sigma \rangle_{\rm ex} =  \int_0^u \sigma_L f_L(1-f_L) d\sigma_L / 4$ for this process.  For comparison, we also calculate the difference of the Shannon entropy between the initial and final distributions of the dot, denoted as $\Delta S$.  Figure 1 (b) shows $\langle \sigma \rangle_{\rm ex}$ (the solid line) and $-\Delta S$ (the dashed line) versus $u$.  They are coincident with each other up to the order of $O(u^2)$, which is consistent with the extended Clausius equality discussed in Sec.~IV.~ B with $u = O(\varepsilon) = O(\Delta)$.
\
\section{Conclusions and Discussions}
We have derived the geometrical expressions of the excess entropy production for quasi-static transitions between NESSs: Eq.~(\ref{berry2}) for $S_{\rm ex}(i \chi)$ and Eq.~(\ref{average_e}) for $\langle \sigma \rangle_{\rm ex}$.  Our results imply that  the vector potentials  $\langle 1_\chi | d |  p_\chi^S \rangle$  and $\langle 1' | d | p^S \rangle$ play  important roles in SST.   We have also derived condition~(\ref{no_curvature}) that a scalar thermodynamic potential exists. 

We note that the arguments in Secs.~II and III are  not restricted to the case of entropy production $\sigma_{xy}^\nu$, but can  be formally applied to an arbitrary quantity $f_{xy}^\nu$ that satisfies $f_{xx}^\nu = 0$.  In fact, even if we replace $\sigma_{xy}^\nu$ by any $f_{xy}^\nu$, the formal expressions of the main results in Sec.~III remain unchanged.  However, we have explicitly used the properties of $\sigma_{xy}^\nu$ such as Eq.~(\ref{entropy}) in Sec.~IV.

We also note that, as is the case for the gauge theory, we can rephrase our results (\ref{berry1}) and (\ref{berry2}) in terms of differential geometry~\cite{Nakahara}.  We consider a trivial vector bundle whose base manifold is parameter space $\{ \bm \alpha \}$.  The fiber is $\mathbb C$, and $c_0 (t)$ in Eq.~(\ref{c_0}) is an element of the fiber.   Then $\langle 1_\chi | d | p_\chi^S \rangle$ is a connection form,  and Eq.~(\ref{equation}) describes the parallel displacement of $c_0$  with the connection along curve $C$.

In this paper, we have assumed that nonequilibrium dynamics is modeled by a Markovian jump process with transition rate $R$ being diagonalizable.  To generalize our results to other models of nonequilibrium dynamics is a future issue.  For example, it is worth investigating whether our result can be generalized to Langevin systems.
Moreover, to investigate the usefulness of our results in nonequilibrium thermodynamics is also a future challenge.


\begin{acknowledgments}
We are grateful to T. S. Komatsu, N. Nakagawa, T. Nemoto, J. Ohkubo, S. Okazawa, K. Saito, S. Sasa, and H. Tasaki for valuable discussions.  This work was  supported by the Global COE Program ``The Next Generation of Physics, Spun from Universality and Emergence'' from the Ministry of Education, Culture, Sports, Science and Technology (MEXT) of Japan,  the Grantin-Aid of MEXT (Grants No.~21540384),  and the Grant-in-Aid for Research Activity Start-up (Grants No.~11025807).
\end{acknowledgments}


\end{document}